\documentclass[entropy,article,accwpt,moreauthors,pdftex]{Definitions/mdpi} 
\usepackage{comment}
\preto{\abstractkeywords}{\nolinenumbers}
\firstpage{1} 
\pubvolume{xx}
\issuenum{1}
\articlenumber{5}
\pubyear{2019}
\copyrightyear{2019}
\history{Received: date; Accepted: date; Published: date}

\Title{Asymptotic behavior of memristive circuits}

\Author{Francesco Caravelli$^{1,\dagger}$\orcidA{} }

\address{$^{1}$ \quad caravelli@lanl.gov}

\firstnote{Theoretical Division and Center for Nonlinear Studies, Los Alamos National Laboratory, Los Alamos, New Mexico 87545, USA} 

\abstract{The interest in memristors has risen due to their possible application both as memory units and as computational devices in combination with CMOS. This is in part due to their nonlinear dynamics, and a strong dependence on the circuit topology. We provide evidence that also purely memristive circuits can be employed for computational purposes. In the present paper we show that a polynomial Lyapunov function in the memory parameters exists for the case of DC controlled memristors. Such Lyapunov function can be asymptotically approximated with binary variables, and mapped to quadratic combinatorial optimization problems. This also shows a direct parallel between memristive circuits and the Hopfield-Little model.   In the case of Erdos-Renyi random circuits, we show numerically that the distribution of the matrix elements of the projectors can be roughly approximated with a Gaussian distribution, and that it scales with the inverse square root of the number of elements. This provides an approximated but direct connection with the physics of disordered system and, in particular, of mean field spin glasses. Using this and the fact that the interaction is controlled by a projector operator on the loop space of the circuit. We estimate the number of stationary points of the approximate Lyapunov function and provide a scaling formula as an upper bound in terms of the circuit topology only.}

\keyword{Memristive circuits ; Spin models ; disordered systems} 

\begin{document}
\vspace{6pt} 
\section{Introduction}
The study of memristors has become recently an area of interest \cite{indiveri,Avizienis,Stieg12} for a variety of reasons. This is not only due to the fact that memristors are considered by many the fourth circuit element, but also due to their potential applications and their interesting dynamics.
 In particular, it has been understood from a purely computational theory perspective that the combination of memristive and CMOS components leads to universal computing machines, called memcomputers \cite{traversa14b}.
From a theoretical perspective, circuits made of memristors have been shown to exhibit non-trivial dynamics which can in principle serve for computational purposes  \cite{Caravelli2015,Caravelli2016,Caravelli2019,Caravelli20192}.   
 Memristors are passive components which can be thought of as a time varying resistance sensitive to the either the current or the voltage,  which in turn depends on a dynamical internal state variable. The main characteristic of a memristor is a pinched hysteresis loop in the Current-Voltage diagram when controlled in alternate voltage \cite{chua76a, strukov08, stru12}. 
The aim of this paper is characterizing the asymptotic behavior of purely memristive circuits (i.e. only memristors) for general circuit topology. 
An open question which has not insofar been answered at a theoretical level is what is the role of the circuit topology in the relaxation process of memristive circuits. It is in fact thought that the topology plays a role in the optimization capabilities of these circuits with memory (also called information overhead by Di Ventra and Traversa \cite{traversa14b}). We provide a precise answer for the simpler class of circuits in which we know the exact role of topology, and show that the dynamical interactions across the circuit map into a coupling term between the memristors for the Lyapunov function for the dynamics.
Specifically, we show that memristive circuits, if controlled with constant voltage, can perform naturally a local unconstrained optimization, specifically a Quadratically Unconstrained Binary Optimization (QUBO) \cite{QUBO}.
In order to gain some theoretical understanding on the complexity of the asymptotic states we employ the Kac-Rice formula (average number of stationary points) to provide a rough upper bound on the complexity of the function. In the process of studying random circuits, we also show the connection between memristor dynamics on random circuits and the Sherrington-Kirkpatrick model.

Our results not only establish a direct and analytical connection between (local) optimization and memristors, but potentially introduces a new class of (heuristic) optimization algorithms for combinatorial problems based on the first order dynamical equation studied in the present paper.
In fact, an interesting byproduct of the present paper is that the Lyapunov function we derived goes beyond the physically implementable circuits, e.g. in principle it applies to cases in which the projection operator is replaced with (semi-)positive operators. As an application, we map a classical problem of stock returns optimization (Markowitz) into ours, and test the performance of the found optimization procedure. We provide an instance of optimization of the Nikkei 225 dataset in the Markowitz framework, and show that it is competitive at least compared to exponential annealing, which usually performs poorly on hard combinatorial problems.

\section*{Memristive circuits}
Before we introduce the dynamics for a generic circuit, we first briefly discuss the type of memristors under scrutiny \cite{strukov08}.
Specifically, we consider memristors whose internal dynamics (the parameter $w$) depends on the current only, and in which the resistance depends linearly on an internal parameter $w$, and satisfy the linear relation $R(w)= R_{on} (1-w)+R_{off} w,$ with $0\leq w(t)\leq1$. We consider in particular the time evolution of a single memristor with diffusive dynamics \cite{AtomicSwitch1,AtomicSwitch2} and for an applied voltage $S$:
\begin{equation}
\frac{d}{dt} w(t)=\alpha w(t)-\frac{R_{on}}{ \beta } I = \alpha w(t)-\frac{ R_{on}}{ \beta } \frac{S}{R(w)}
\label{eq:singlemem}
\end{equation}
which shows that a competition between drift and decay occurs. This type of memristors has been considered for machine learning applications.  With our convention,  $R_{on}$ and $R_{off}$ are the limiting resistances for $w=0$ and $w=1$ respectively ($R_{off}>R_{on}>0$), and $\alpha$ and $\beta$ are constants which set the timescales for the relaxation and excitation of the memristor respectively.\footnote{It is worth mentioning that our model is different from the one introduced in \cite{strukov08} as it has not only opposite polarity, but also a decay function which is slightly different.}
In a recent paper \cite{mean-field} it has been observed that mean field theory and techniques from statistical physics can be applied to study a specific circuit topology. In that paper, it was also noted that a Lyapunov function exists, and that zero temperature mean field theory provides a good estimate for the average asymptotic dynamics for a single mesh of memristors. In the present paper we extend some of these results to a more general class of memristive circuits, which sheds new light onto this type of nonlinear systems. Specifically, in \cite{Caravelli2015} the following differential equation was derived for a generic but purely memristive circuit:
\begin{eqnarray}
 \frac{d \vec W}{d t} &=&\alpha \vec W- \frac{1}{\beta}\mathcal ( I+\xi\ \Omega  W)^{-1}  \Omega  \vec { S}( t),
\label{eq:dynnn5}
\end{eqnarray}
where $\vec S$ is the vector of voltage sources in series to each memristor, while $I$ is the identity matrix.  It is immediate to observe that the nonlinearity is controlled by the parameter $\xi=\frac{R_{off}-R_{on}}{R_{on}}$. It is important to note that $\Omega$ and $W$ are matrices, while $I$ is the identity matrix. $W$ is the diagonal matrix with the memristor memory values $w_i(t)$ on the diagonal, meanwhile $\Omega=A^t(A A^t)^{-1} A$ is a projector operator on the fundamental loops of the circuit, being $A$ the cycle matrix of the directed graph representing the circuit flows\footnote{The matrix $A$ has only $-1,+1,0$ and zero values. Given an orientation of each loop and edge, then $A_{i\beta}$ has dimensionality $L\times N$, where $N$ is the number of memristors and $L$ the number of fundamental loops. If a memristor $\beta$ has the same orientation of a loop $i$, then $A_{\beta i}=1$, and $-1$ if opposite. $A_{\beta i}=0$ if the memristor $\beta$ does not belong to the loop $i$. The matrix $\Omega$ is not diagonal because edges of a graph can belong to more than one loop.}. The richness of the dynamicsl of this equation has been characterized in \cite{Caravelli2015}, while the locality properties of $\Omega$ in \cite{Caravelli2017}.
The important observation that we anticipate is that the dynamical eqn. (\ref{eq:dynnn5}) has at least one Lyapunov function which we have derived analytically.  While the derivation is provided in the Appendix, here we provide only its functional form:
\begin{eqnarray}
L( W)&=&-\frac{\alpha}{2}\sum_i W_i^2-\frac{\alpha\xi}{3}  \sum_{i} \Omega_{ii}  W_i^3\nonumber \\
&-&\alpha\xi  \sum_{i\neq j} \Omega_{ij}  W_i W_j^2+\frac{1}{\beta} \sum_{ij} W_i \Omega_{ij} S_j,
\label{eq:Lyapunovf1}
\end{eqnarray}
from which a few key facts can be immediately observed. For instance, the role of $\Omega$ is the one of a coupling matrix, as one would expect, and that it is not quadratic in the internal memory variables. Eqn. \ref{eq:Lyapunovf1} satisfies $\frac{d}{dt} L(W)\leq0$ whenever $M_i=-2 \alpha \xi \sum_{j\neq i} W_j \Omega_{ji} W_i$ is small (thus for very weakly interacting memristors, as it depends only on the offdiagonal terms) and $\frac{d}{dt} L(W)=0 \leftrightarrow \frac{d}{dt} \vec W=0$ (see the details in the Appendix). One can obtain a more precise bound on the derivative of the Lyapunov function. Let us define
$|| \Omega \vec S||^2= N^2 s^2(N)$. We have proved in the Appendix that if
\begin{equation}
4 \xi^2(1+\xi) \bar \Omega^2 < \frac{s(N)^2}{\alpha^2 \beta^2}-2\frac{s(N)}{\alpha \beta},
\label{eq:phasediagram}
\end{equation}
then $\frac{d}{d} L(W)<0$, where $\bar \Omega=\max_{ij}|\Omega_{ij}|$. It is interesting to note that the order parameter $\frac{s(N)}{\alpha \beta}$ is a generalization of the one found in \cite{mean-field} and which controlled the asymptotic state of the effective mean field circuit. Here again it plays a role. Note that $\xi^2(1+\xi) \bar \Omega^2$ is what controls the height of the parabola, and intuitively the larger then nonlinearity the higher the voltage has to be for the system to relax according to the Lyapunov function. Thus, eqn. (\ref{eq:phasediagram}) establishes a dynamical phase diagram.

\begin{figure}
\centering
\includegraphics[scale=0.7]{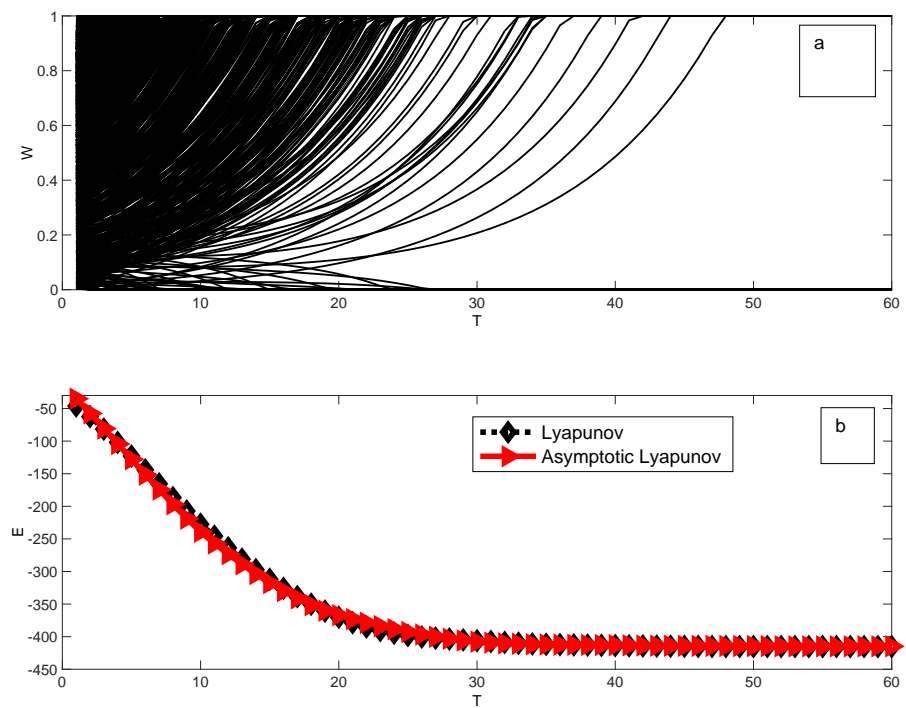}
\caption{Dynamics of the memory is shown in Fig. (a) and the corresponding evolution of the Lyapunov function of eqn. (\ref{eq:Lyapunovf}) (b, continuous line) for 8750 memristors and for the asymptotic function of eqn. (\ref{eq:Lyapunovfa}) Fig. (b-dashed line). We considered $\alpha=0.1$, $\beta=1$ and $\xi=10$, for $\Omega_{ij}$ from a random circuit of the Erdos-Renyi type with $p=0.9$. The sources $\vec S$'s elements were chosen at random between $[-0.05,0.05]$.}
\label{fig:dynamics}
\end{figure}

\begin{figure}
\centering
\includegraphics[scale=0.7]{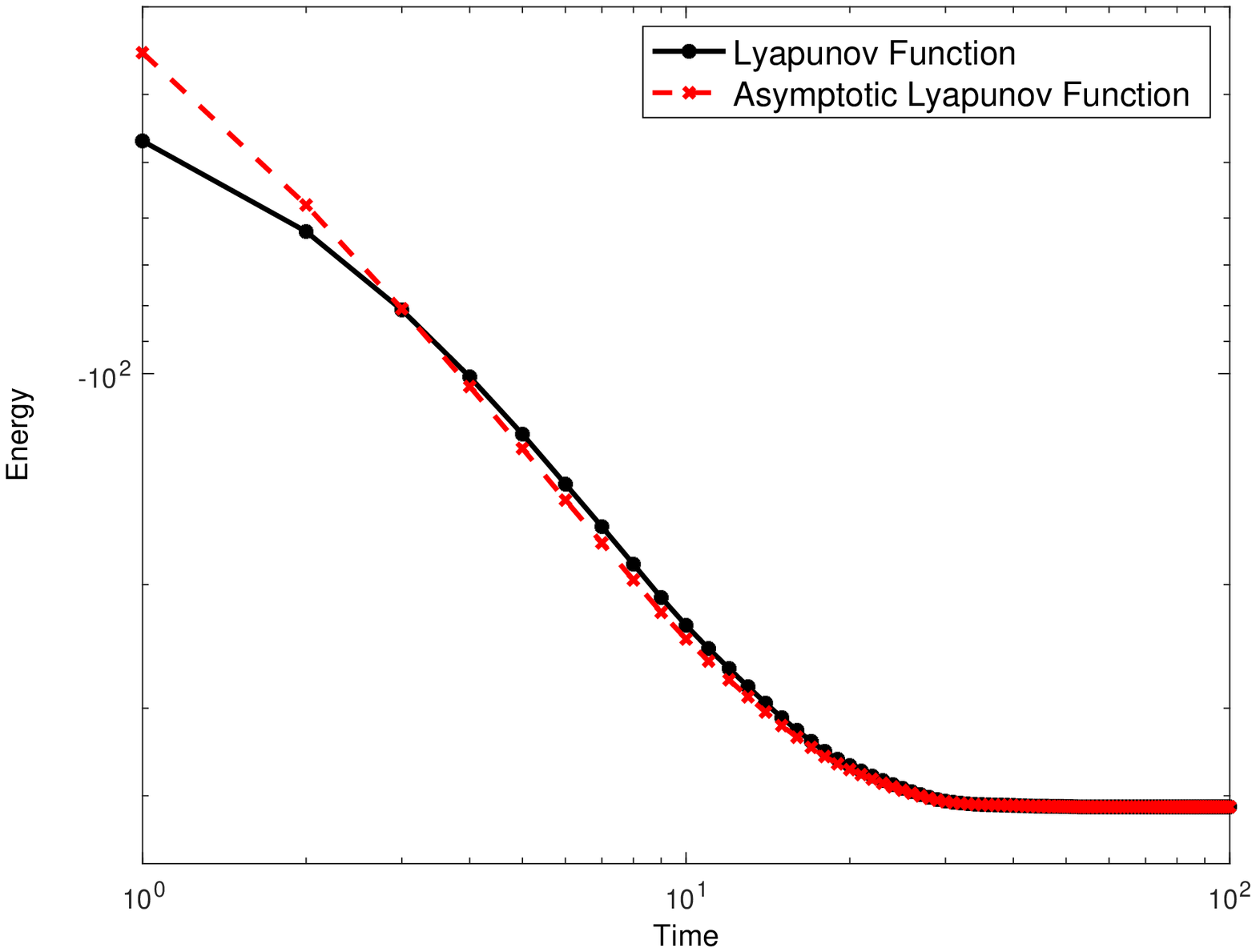}
\caption{Dynamics of the circuit (a) and corresponding evolution of the Lyapunov function (b).}
\label{fig:dynamics2}
\end{figure}
We stress that the dynamics is constrained in the hypercube $[0,1]^N$, where $N$ is the number of memristors, and thus the Lyapunov function above when a memristor reaches the boundaries of this domain. However, if we now use the fact that the dynamics is controlled with constant voltage each memristor will eventually reach the asymptotic value 1 or 0. In this asymptotic limit, we note that the function above can be interpreted as a spin-like model asymptotically.\footnote{In \cite{mean-field} this phenomenon has been partly explained by the emergence of unstable fixed points in the dynamics (although only in the mean field approximation). For the specific case of the eqn. (\ref{eq:dynnn5}) however, it is necessary to go beyond this approximation, for which each memristors will have (intuitively) a different fixed point associated to it. This phenomenon, which is circuit dependent, is associated to the structure of the couplings $\Omega_{ij}$, and will be studied elsewhere.} In this case, the Lyapunov function reads
\begin{eqnarray}
L( W\left(t\gg1\right))&\approx& -[\frac{\alpha}{2}\sum_i W_i+\frac{\alpha\xi}{3}  \sum_{i} \Omega_{ii}  W_i \nonumber \\
&+&\alpha\xi  \sum_{i\neq j} \Omega_{ij}  W_i W_j-\frac{1}{\beta} \sum_{ij} W_i \Omega_{ij} S_j ] \nonumber \\
\equiv L_a(W)&=& -[\sum_i W_i \left(\frac{\alpha}{2} +\frac{\alpha\xi}{3}\Omega_{ii} -\frac{1}{\beta} \sum_j \Omega_{ij} S_j\right) \nonumber \\
&+&\alpha\xi  \sum_{i\neq j} \Omega_{ij}  W_i W_j ],
\label{eq:Lyapunovfa}
\end{eqnarray}
where we have used the fact that $W_i^n=W_i$ if $W_i=1,0$. An effective external field $h_i=\frac{\alpha}{2} +\frac{\alpha\xi}{3}\Omega_{ii} -\frac{1}{\beta} \Omega_{ij} S_j$ emerges in this asymptotic approximation.
We also test numerically both the Lyapunov function of eqn. (\ref{eq:Lyapunovf}) and eqn. (\ref{eq:Lyapunovfa}) in Fig. \ref{fig:dynamics}. Meanwhile in Fig. \ref{fig:dynamics} (a) we plot the evolution of each memory element, in Fig. \ref{fig:dynamics} (b) we plot the evolution of both the Lyapunov function and the asymptotic approximation of it, which remains close to the exact one. On the other hand, in Fig. \ref{fig:dynamics2} we show the difference between the obtained Lyapunov function and the asymptotic one as a function of time.
This mapping is reminiscent of the case of continuous neuronal networks introduced by Little \cite{Little} and then Hopfield in a series of important papers \cite{Hopfield1,Hopfield2}. The Little-Hopfield model has sparked interest from the Statistical Physics community since the very beginning \cite{SGHopfield}.
In the past years this particular line of study has been subject of scrutiny by many experimental groups \cite{HopMemr1,HopMemr2,HopMemr3}. The key difference is that in each case, these were studied in conjunction with ordinary and active (CMOS) electronic components to build Hopfield learning networks. We argue instead that memristive circuits \textit{per se} form a special kind of Hopfield network defined by the Lyapunov function above, without the need of extra components.


\subsection*{Functional complexity for random circuits}
We now study random circuits, in particular for the function complexity of eqn. (\ref{eq:Lyapunovfa}). First, this will allow us to show an interesting connection to the field of disordered systems, and in this approximation provide estimates of the complexity of the Lyapunov function. In fact, minimizing quadratic function with discrete variables is in general a hard problem to solve. We are however in the situation in which the problem at hand is in nature both continuous in time and in the variables (being these between $0$ and $1$), but naturally provides an answer for the more complicated case usually associated to finding the ground state of a frustrated spin systems. Despite the fact that we are not able in the present paper to provide a complete answer to which optimization class a system of purely memristive circuits belongs to, we try to provide some answers to the questions above using some techniques introduced to study random polynomials \cite{rgf,fyodorov}. This will be important in light of the fact that we have shown that memristors perform a local optimization, rather than a global one, and that the circuit constraints should enter somehow.

The number of stationary points of a generic multi-varied function $L(W)$ can be estimated by:
\begin{equation}
 \#  =\sum_{\{\vec w=1,0\}}  \text{det}(\partial^2_{\partial w_i \partial w_j} L(\vec w)) \prod_i \delta\left(\partial_{w_i} L(\vec w)\right).
 \label{eq:sp}
\end{equation}
Thus, if we aim to consider the expected number of stationary points $\#$, then it does make sense to consider the (quenched) average of the quantity in eqn. (\ref{eq:sp}) for the distribution for $\langle\cdot \rangle_{P(\Omega)}$ with respect to the coupling matrix $\Omega$. One important observation is that because $\Omega$ is a projector, $\text{det}(\partial_{W_i} \partial_{W_j} L(W))$ can be calculated exactly without knowledge of the elements distribution, but only using the fact that $\Omega$ is a projector (details in the Appendix).  
We find that the average number of stationary points can be split in the form:
\begin{equation}
\langle \# \rangle=(\sqrt{3} \alpha \xi)^N \left( 1-\frac{1}{\sqrt{N}}\right)^L Z(\vec S,Q)
\label{eq:scalingp}
\end{equation}
where $L$ is the number of fundamental loops of the circuit, and where:
\begin{equation}
Z(\vec S,P(\Omega)) =\langle\sum_{\{\vec w=1,0\}} \prod_i \delta\left(\partial_{w_i} L(\vec w)\right)  \rangle_{P(\Omega)}.
\end{equation}
depends on the distribution of the elements of the matrix $\Omega$. 

\begin{figure}[t!]
\centering
\includegraphics[scale=0.4]{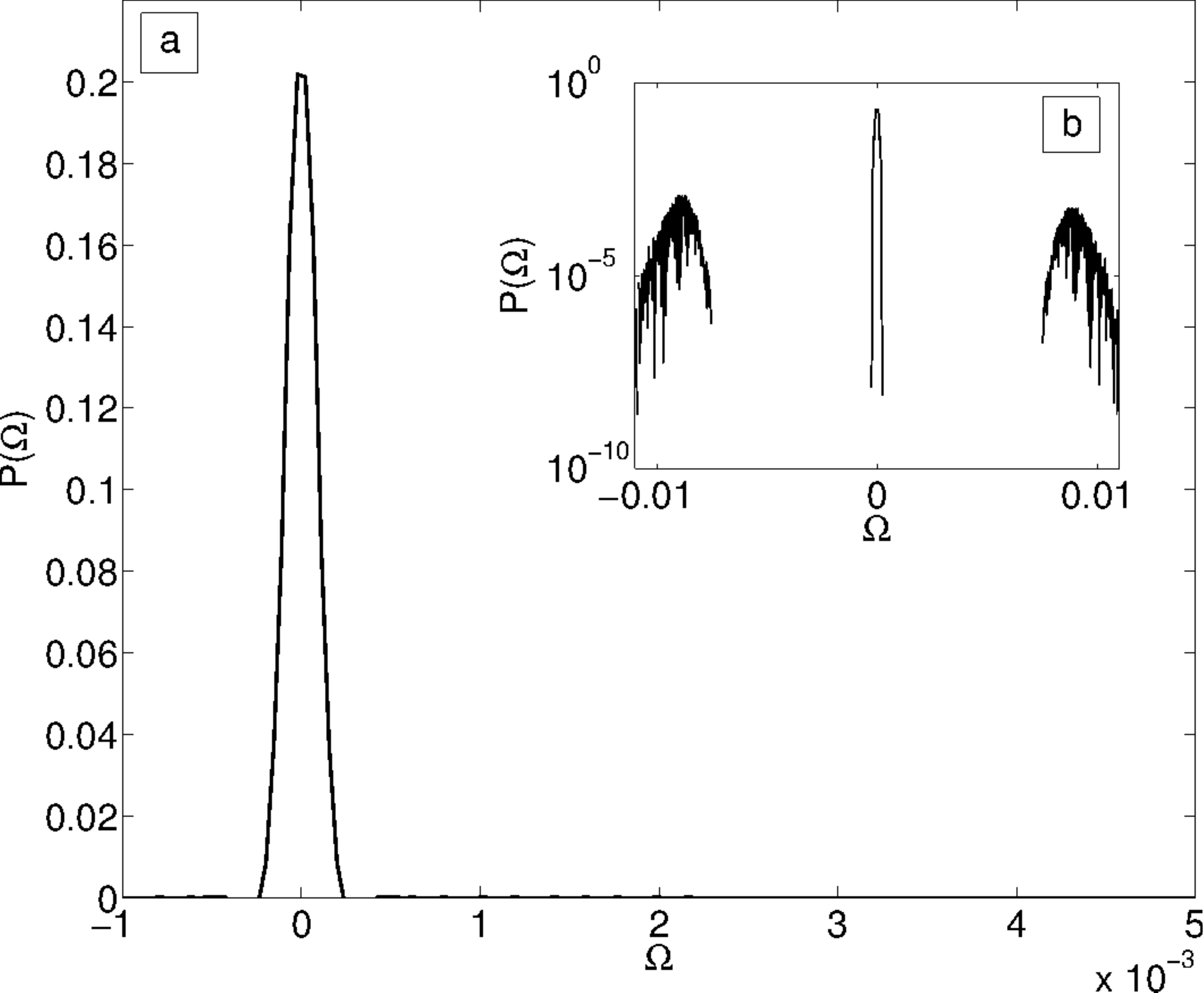}
\caption{Distribution of the elements of the matrix $\Omega$ in the case of an Erdos-Renyi graph.}
\label{fig:distomega}
\end{figure}

We first need, then,  to understand the distribution $P(\Omega)$ for random circuits. Since the analytical calculation of such distribution goes beyond the scope of this paper, we perform such analysis numerically. We first generate random circuits with Erdos-Renyi graph above the percolation threshold ($p=0.7$). We are interested, in particular, in the scaling with respect to the number of memristors $N$. For fixed $\Omega$, the distribution $P(\Omega)$ of off-diagonal elements is shown in Fig. \ref{fig:distomega} (a). We observe that although this seems to be unimodal, a careful look shows that it is not. We observe in Fig. \ref{fig:distomega} (b) that there is a non-zero probability of having elements not distributed around zero, but that these are orders of magnitudes smaller than the central distribution. While it is not surprising that the distribution is not completely random (thus, there are correlations among elements) since these are constrained by $\Omega^2=\Omega$, it is surprising to note that we can roughly approximate this distribution with a unimodal one.
We are in particular interested in how the width of the distribution scales with the number of memristors. This can be calculated exactly if we know the scaling of the diagonal elements, as we will see below. In Fig. \ref{fig:scaling} we show how the diagonal elements of $\Omega$ scale with $N$. We fit first the diagonal elements, precisely $1-\Omega_{ii}$, showing that it scales as a power law to a good approximation. If we perform a fit using the functional form $\Omega_{ii}\approx 1-\frac{c}{N^\alpha}$, we obtain the best fit values  $c=1.74557\approx \sqrt{3}$ and find that $\alpha=0.5043\approx1/2$. We will, since now on, consider these values in what follows.
It is important to note that the scaling above is sufficient to obtain a scaling for the off-diagonal elements as well. We define the matrix of off-diagonal element $G$, as $\Omega\approx (1-\sqrt{\frac{3}{N}})I+G$. Since we have that $\Omega^2=\Omega$, it is easy to see that $G$ must satisfy the equation $G^2+(1-2\sqrt{\frac{3}{N}}) G+ \sqrt{\frac{3}{N}}(\sqrt{\frac{3}{N}}-1)I=0$. For $N\gg1$ this implies that also $G$ satisfies the equation $G^2=G$ as well. We can at this point solve for an eigenvalue equation for $G$, and obtain $\lambda=\{\sqrt{\frac{3}{N}},-1+\sqrt{\frac{3}{N}}\}$. Since this must be a projector for large values of $N$, we must have $\Omega\approx I+\sqrt{\frac{3}{N}} Q$, where $G=\sqrt{\frac{3}{N}} Q$, with $Q$ again a projector. Thus, the scaling approximation obtained for random Erdos-Renyi type circuits, we have
$\Omega_{ii} \rightarrow 1$, meanwhile $\Omega_{ij}\approx \frac{\sqrt{3}}{\sqrt{N}} Q_{ij}$. This is the scaling we will use in the following. 

\begin{figure}[ht!]
\centering
\includegraphics[scale=0.4]{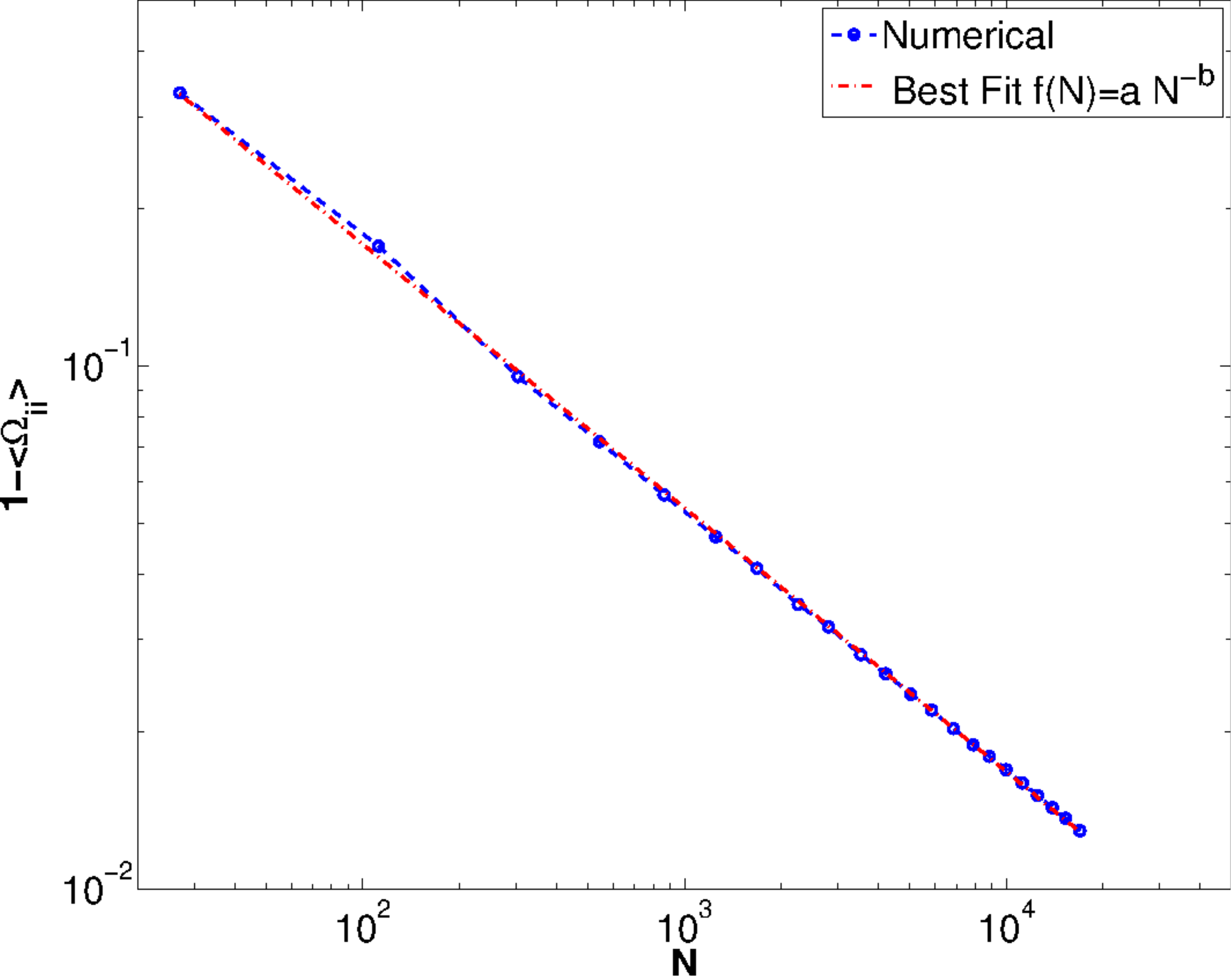}
\caption{Scaling of the components $\Omega_{ii}$ as a function of the number of memristors for randomly generated circuits}
\label{fig:scaling}
\end{figure}

While in what follows we show only the intermediate results, all the exact calculations are provided in the Appendix. We can in fact a this point calculate  $Z$ in eqn. (\ref{eq:sp}) for random circuits. We  assume that $P(Q_{ij})=\frac{1}{\sqrt{2\pi \sigma^2}} e^{-\frac{Q_{ij}^2}{2 \sigma^2}}$, where $\sigma$ is an effective width for $N=1$. In this case, calculating $Z$ can be done by means of Gaussian integrals.  In this case, we obtain, in the limit $\alpha\xi\gg1$:
\begin{equation}
\langle \# \rangle\approx \left(1-\frac{1}{\sqrt{N}}\right)^L (\frac{ \sqrt{3} }{\sqrt{\pi}\sigma} )^N.
\end{equation}
We now use Euler's relation $L=N- V + \chi$, where $V$ is the number of vertices, $\chi$ the Euler characteristic and use the Erdos-Renyi relation $N=p(V) \frac{V(V-1)}{2}$ where $p(V)$ is the edge probability. Now we have that, since $\chi>0$ and only topological, while $L$ scales linearly for large $N$ in the number of memristors. We now observe that given $\rho=\frac{\sqrt{3}}{\sqrt{\pi}\sigma}$, we have that in the limit $N\rightarrow\infty$, $\langle \# \rangle$ goes to infinity for $\rho> 1$, while it goes to zero for $\rho\leq 1$, as $\lim_{N\rightarrow \infty} (1-\frac{1}{\sqrt{N}})^N=0$. We thus obtain the result that for $\sigma>\frac{\sqrt{3}}{\sqrt{\pi}}$ the Lyapunov function seems to have a large number of stationary points, and thus a hint of the hardness of the minimization problem.

It is easy at this point to provide an approximate mapping between the asymptotic Lyapunov function and the Hamiltonian of a mean-field spin glass of the Sherrington-Kirkpatrick type with an external field:
\begin{equation}
L(\sigma)=\sum_i \tilde h_i \sigma_i + \frac{\alpha \xi \sqrt{3}}{\sqrt{N}} \sum_{i\neq j} Q_{ij} \sigma_i \sigma_j
\label{eq:glassy}
\end{equation}
where we used the mapping $W_i=\frac{1}{2}\sigma_i-\frac{1}{2}$ with $\sigma_i=\pm1$, and $
\tilde h_i=\frac{h_i}{2}- \frac{1}{2}\sum_j Q_{ij}$, and disregarded an unimportant constant which only shifts the function.
The result above is a generalization of what has been found in \cite{Caravelli2016ml} in the limit $\xi\ll 1$, e.g. the fact that the dynamics of memristors follows a constrained gradient descent (Rosen projections). Moreover, it provides a physical system to test experimentally the physics of mean field spin glasses \cite{parisi}.

\begin{figure}
\centering
\includegraphics[scale=0.4]{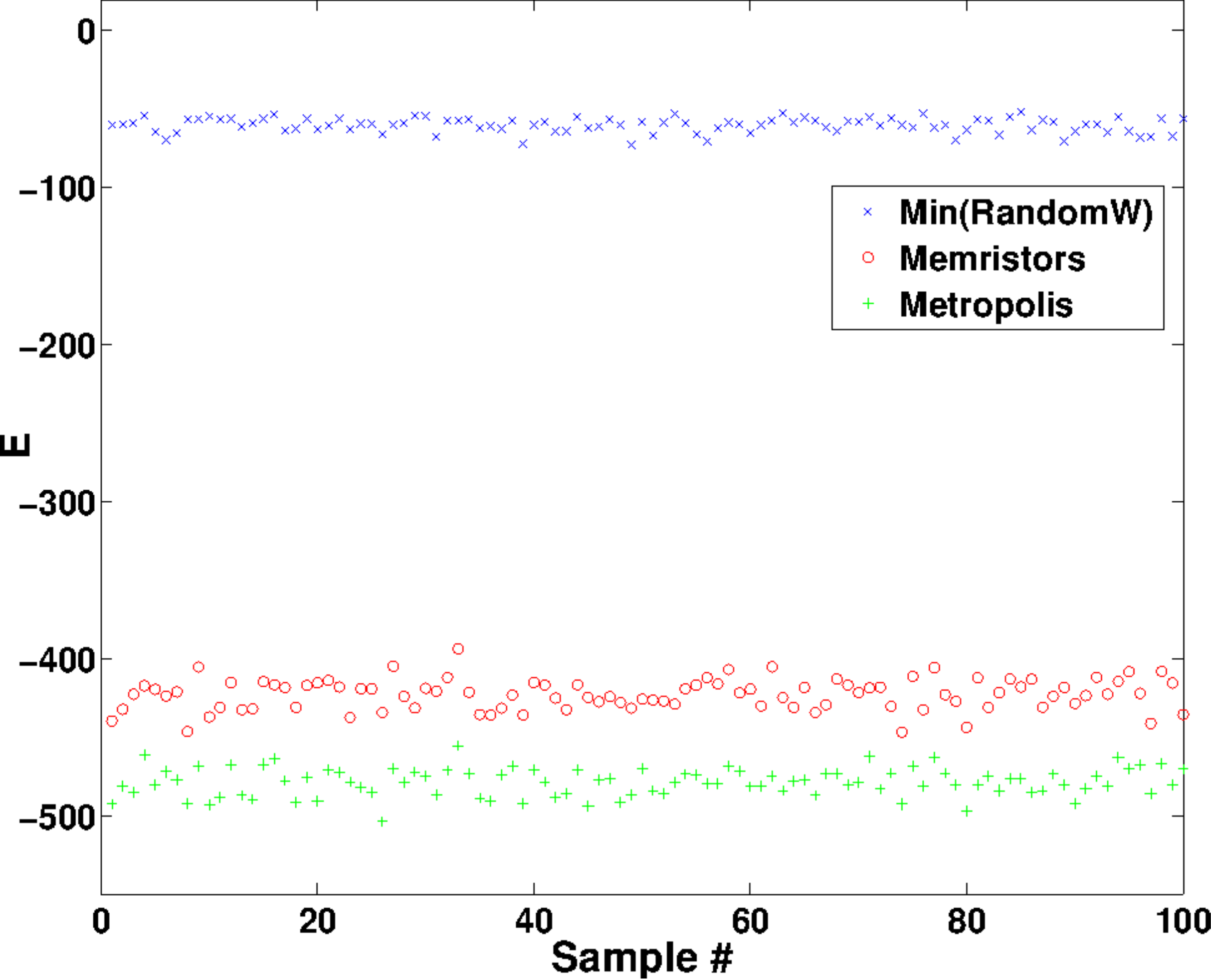}
\caption{Obtained minimum energy using the minimum over 100 random binary values (blue crosses), roghly 8000 steps of exponential annealing (green pluses) and 60 time steps using the memristor dynamics (red circles). We generated each sample with approximately $800-900$ memristors, fixing the percolation parameter to $p=0.7$, with $\xi=10$, $\alpha=0.1$ and $\beta=1$. The numerical integration was performed using a simple Euler-Newton integration with integration step $dt=0.1$, and total number of steps $T=1000$. We truncated the Metropolis annealing to $10*N$ time steps, with $N$ the number of memristors, and with an annealing rate $\lambda=0.995$, and an exponential annealing law for the temperature given by $Temp(k)=100*\lambda^k$.}
\label{fig:minimum}
\end{figure}

\section{Asymptotic state recollection and combinatorial optimization}
\begin{figure}
\centering
\includegraphics[scale=0.3]{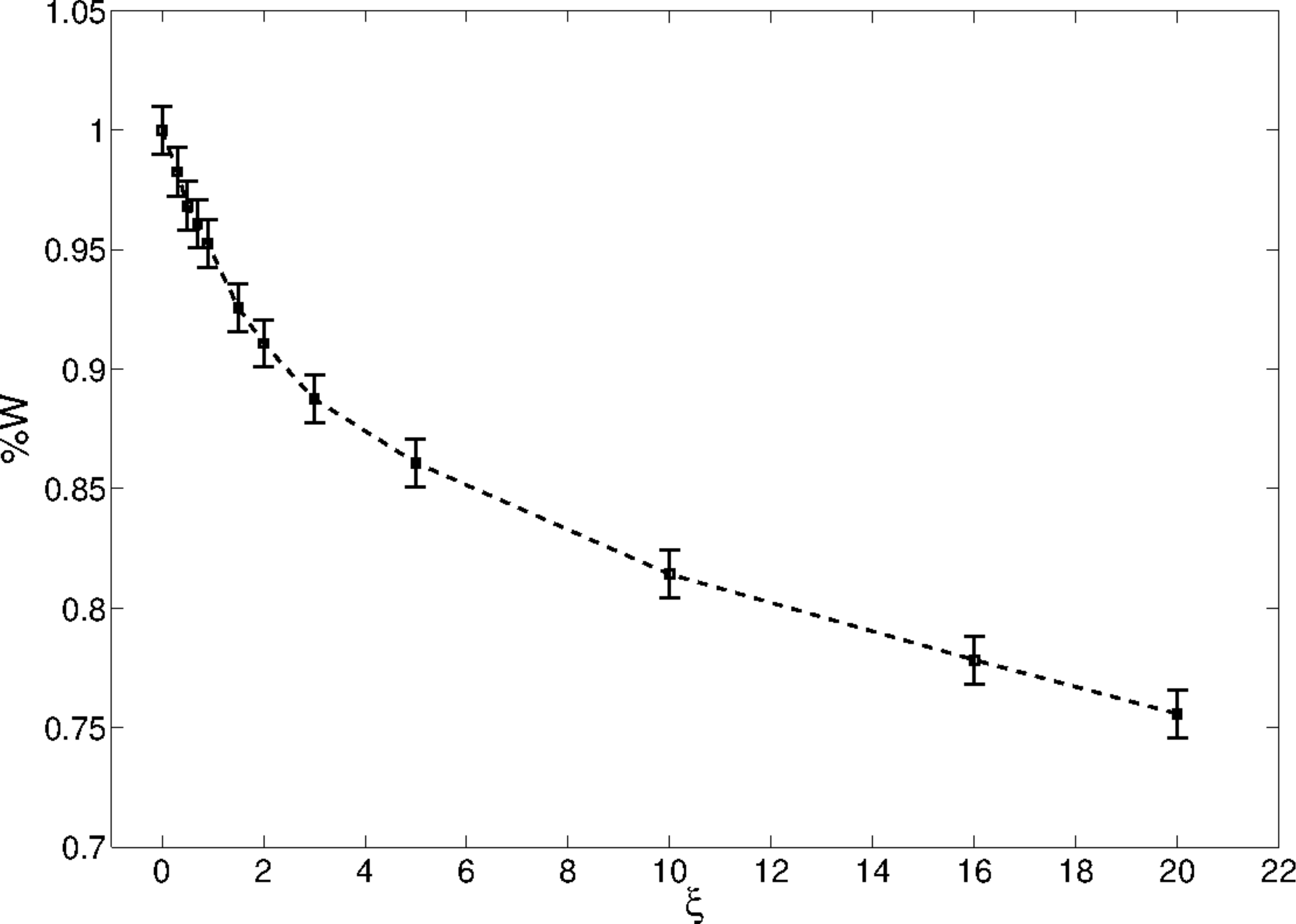}
\caption{Percentage of exactly predicted values of memristors according to eqn. (\ref{eq:percpred}) as a function of $\xi$. The error bar are calculate from 100 samples, with approximately 800 memristors (on average) and $\xi=10$, $\alpha=0.1$ and $\beta=1$, with the vector $S$ drawn at random between $[-0.5,5]$.}
\label{fig:expred}
\end{figure}

One important question for memristive circuits is what is the asymptotic state they reach. As we have argued, this can be answered by looking at the minima of the Lyapunov function for the case of constant voltages. Although this is a hard problem to solve in its full generality, we can use some approximate analytical formulae to provide an answer in at least for some region of the parameters. First, we consider the case  $\alpha=0$. It is easy to see that we can integrate the equation \cite{Caravelli2016ml} to obtain:
\begin{equation}
\vec W+\frac{\xi}{2} \Omega \vec W^2=-\frac{1}{\beta}\Omega \vec S (t-t_0)+\vec c
\label{eq:unim}
\end{equation}
where $c$ is an integration constant. Eqn. (\ref{eq:unim}) is a \textit{unimodular quadratic vector equation},
which is the special case of vector equations of the form $\vec x+\vec b(\vec x,\vec x)=\vec c$. This equation does not have an exact solution in closed form, but several numerical methods have been developed \cite{poloni}. Here we use a heuristic method to identify what are the asymptotic values of $\vec W$. Again we use the fact that in the limit $t\rightarrow \infty$, we observe numerically the asymptotic values of $W$ are either $0$ or $1$. 
For $\xi=0$, we have $\vec W(t=\infty)=\frac{1-\text{sign}(\Omega \vec S)}{2}$. For $\xi>0$ and in the asymptotic limit, $\lim_{t\rightarrow\infty} \vec W^2\approx \vec W$. Using this approximation, we obtain:
\begin{equation}
(I+\frac{\xi}{2} \Omega)\vec W=-\frac{1}{\beta}\Omega \vec S (t-t_0)+\vec c.
\label{eq:unim}
\end{equation}
We then can obtain the asymptotic formula:
\begin{eqnarray}
\vec W(t=\infty)&\approx&\frac{1-\text{sign}((I+\frac{\xi}{2} \Omega)^{-1}\Omega \vec S)}{2} =\frac{1-\text{sign}( \Omega \vec S)}{2}, \nonumber \\
\label{eq:percpred}
\end{eqnarray}
where we note that in the last equation we have used the fact that $1+\xi/2>0$ if $0<R_{on}<R_{off}$. 
However, we note that the exactness of the asymptotic behavior depends only on $\xi$ and not on $\alpha$. 
The results on how well eqn. (\ref{eq:percpred}) predicts the asymptotic behavior of the circuit as a function of $\xi$ are shown in Fig. \ref{fig:expred}. Despite our heuristic method and approximation, the obtained values are a good approximation for small values of $\xi$ of the real system. Using the formula above, we now provide the connection with the state recollection of neural networks. We note that the projection operator $\Omega$ can be written as \cite{Caravelli2017} $
\Omega_{ij}=\sum_{l=1}^L \tilde A_{i}^l\tilde A_{j}^{l}$,
where $\tilde A_i^l$ is the orthonormalized loop matrix of the circuit. From the theory of neural networks we are aware that the number of stored patterns equals the number of independent eigenvectors of the interaction matrix $\Omega$, which is purely topological, as argued in \cite{Caravelli2016ml}.
In the case of a purely memristive circuit the number of independent memory units number is constrained by the topology of the circuit, and it depends on the number of fundamental loops \cite{Caravelli2016ml}. As we have noted before, the asymptotic behavior of the circuit can be approximated by (where we reinstate $\alpha$)
$$\vec W(t=\infty)\approx\frac{1-\text{sign}(((1-\alpha)I+\frac{\xi}{2} \Omega)^{-1}\Omega \vec S)}{2}, $$
where we realize that the source vector $\vec S$ can act as the recollection mechanism.
If $\vec S=\sum_l \rho_l \tilde A^l$, for a certain proportionality constant $\rho$, then $(\Omega\vec S)_i=\sum_l \rho_l \tilde A^l_i$. Thus, the asymptotic configuration of the memristors will be given by the approximate value:
\begin{eqnarray}
\vec W(t=\infty)&\approx&\frac{1-\text{sign}(((1-\alpha)I+\frac{\xi\rho}{2}  \tilde A^l)^{-1})}{2}\nonumber \\
&=&\frac{1-\text{sign}\left((1-\alpha)I+\frac{\xi\rho}{2}  \tilde A^l\right)}{2},
\end{eqnarray}
which shows that stored patterns are indeed in the loop basis.
The analysis above is reminiscent of the asymptotic state recollection in the Hopfield-Little model. 

We now ask the converse question of how close are the asymptotic states to minima obtained from the Lyapunov functional. As we have seen in the previous sections memristors can be used to provide solutions (possibly sub-optimal) to hard combinatorial problems. This claim should be taken, if not with skepticism, with some care. First we note that $L(W)$ is not positive definite (although it is bounded), and thus we have an asymptotically local stable equilibrium rather than a global one.
We are interested in providing some benchmarks on how well memristors find the minimum of the function $L_a(W)$ compared to other optimization techniques. Given $100$ samples of random circuits based on Erdos-Renyi random graphs with $p=0.7$, in Fig. \ref{fig:minimum} we show the energy attained using memristors in $60$ time steps (red dots), using a Metropolis algorithm with exponential annealing (blue) with over 8000 steps, and using the minimum of 100 random configurations as a reference (blue dots). We see that despite the fact that memristors do not seem to reach the absolute minimum, the energy is closer to the Metropolis result than to the random configuration one. Although it is easy to see that the dynamical equations do not reach the Metropolis attained minimum,  it did take less than 100 iterations for the memristive system to converge using a simple Euler-Newton first order integration. This already shows that, as anticipated, memristors perform a local optimization in this regime. In the next section, however, we apply the algorithm to a specific dataset, where this optimization has some advantage over the exponential annealing.

\section*{An application: minimization of a function}
We have argued that memristors' dynamics can be thought, when these are connected in a circuit, as reaching an asymptotic state which is among the minima of a certain Lyapunov function. 
Albeit real memristors can have a dynamics which is far from ideal, we would like to consider if these could in principle be used for any sort of purpose which goes beyond electronics.
In fact, note that (\ref{eq:dynnn5}) can be used as a quick meta-heuristic method to search for the minimum of a function, or to be given as input to other more complicated and efficient optimization techniques. In realistic circuits, the quadratic form matrix $\Omega$ which occurs in the Lyapunov function is a projector on the cycle space of the graph. Thus, problems for which analog memristive circuits can be used  are naturally those that can be embedded in the cycle basis of a graph \cite{cycle}. However, in principle we can simulate the system also for matrices $\Omega$ that are not necessarily projectors, as we have shown and as we discuss further below. We can thus perform a preliminary test of the applicability of analog memristive circuits in a problem of optimization, and see how the circuit performs compared (for instance) to simulated annealing. 

We thus consider a simple application of practical importance, e.g. studying the problem of investment in a set of assets. For this purpose, we use the 225 Nikkei dataset which is used for benchmarking heuristic optimization algorithms \cite{heur}, available at \cite{heur2}. We can use in fact a memristor-inspired optimization scheme, taking advantage of the fact that it works beyond projector operators $\Omega$.  We mention that the proof of the Lyapunov function relies only on the matrix $I+\xi \frac{\Omega W+W\Omega}{2}$ being positive definite, and thus applies also to any positive matrix and to some non-positive ones (see SI for details) for $\xi$ small enough. Although we cannot use these in a hardware circuit, we can solve the differential equation numerically to infer a minimum heuristically. Using this idea, we ask in which of the assets of portfolio we should invest in (a yes-no answer). The dataset is composed of 225 assets, including returns and the covariance matrix, and can thus use the combinatorial Markowitz functional \cite{Markowitz}.
If we use the fact that maximizing a quadratic function is equivalent to minimizing the function with a minus sign in front, we can write an equality between the Markowitz function for binary variables and the memristive one: 
\begin{eqnarray}
M(W)&=&\sum_i \left(r_i-\frac{p}{2}\Sigma_{ii} \right)W_i-\frac{p}{2} \sum_{i\neq j}  W_i \Sigma_{ij}  W_j \nonumber \\
-L(W)&=& \sum_i W_i \left(\frac{\alpha}{2} +\frac{\alpha\xi}{3}\Omega_{ii} -\frac{1}{\beta} \sum_j \Omega_{ij} S_j\right) \nonumber \\
&-&\alpha\xi  \sum_{i\neq j} \Omega_{ij}  W_i W_j 
\end{eqnarray}
where $p$ is a trade-off parameter, $\vec r$ are the returns and $\Sigma$ is the covariance matrix. From imposing the equality between the two functionals, we observe the necessity of imposing $\Omega=\Sigma$, and thus
\begin{eqnarray}
\frac{\alpha}{2} +\frac{\alpha\xi}{3}\Omega_{ii} -\frac{1}{\beta} \sum_j \Omega_{ij} S_j &=& r_i-\frac{p}{2}\Sigma_{ii},\ \ \frac{p}{2}=\alpha \xi.
\end{eqnarray}
We can thus obtain the source vector to try to force the system towards the right minimum:
\begin{eqnarray}
\vec S=\beta \Sigma^{-1}\left(\frac{\alpha}{2}+(\frac{p}{2}+\frac{\alpha \xi}{3}) \vec \eta-\vec r\right)
\end{eqnarray}
where $ (\eta)_i=\Sigma_{ii}$, e.g. the variance of each return. 
We have the freedom to choose arbitrarily $\beta$, but either $\xi$ or $\alpha$ would be fixed. We observe that we need to invert the covariance matrix, which is a slow but polynomial in the number of variables. Moreover, we see that the harder the problem to solve (the norm of the inverse of $\Sigma$), the smaller $\beta$ has to be chosen, from which we infer the slowness of the process. 

\begin{figure}
\centering
\includegraphics[scale=0.3]{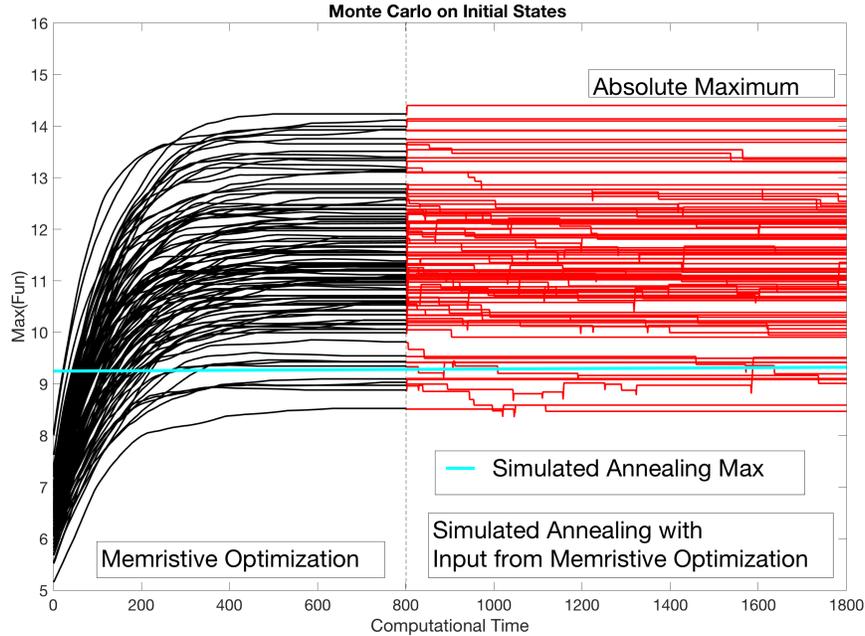}
\caption{Maximization of the return for the Nikkei 225 dataset using a Monte Carlo procedure. We compare the results between the best of 100 Simulated Annealing procedures  (the green line in the plot $\approx 9.15$) versus with the memristive optimization proposed in this paper (black lines), with identical initial conditions. The simulated annealing was conducted with an initial temperature $T=100$, and an exponential annealing rate $\lambda=0.999$ (and effective time steps N=500). We have then used the final states of the Memristive Optimization procedure as an input to a simulated annealing procedure with lower temperature $T_0=0.025$ and identical annealing rate, which obtained the absolute maximum of the return.}
\label{fig:Nikkei}
\end{figure}

We are now in the position to test the memristive minimizationi against an exponential annealing process, which is known to perform poorly in the combinatorial setting. The results for the case of the Nikkei dataset are shown in Fig. \ref{fig:Nikkei}, comparing the Metropolis annealing to the case of the computational time and in the final value obtained. We performed three different tests. First, we randomized the initial states, and performed 100 Simulated Annealing procedures \cite{simann}, with initial temperature $T=100$, an exponential annealing rate $\lambda=0.995$ ($T_k=\lambda^k T_0$) and a number of time steps of $N_T=500 *N$, where $N$ was the number of assets $100$. On the Nikkei 225 dataset, the maximum obtained with the simulated annealing was $r_{max}\approx 9.15$. With the identical initial states, the memristive optimization has obtained a maximum of $r_{max}\approx 14.23$. 
We have then used the final state of the memristive optimization output as an input to an annealing procedure, but with a lower initial temperature $T_0=0.025$, thus effectively fast-forwarding the annealing procedure. This combined optimization obtained the absolute maximum, albeit of only less than a percentage better than the memristive optimization, of $r_{max}\approx 14.40$.

\section*{Concluding remarks}
In the present paper we have analyzed the asymptotic dynamics of memristive circuits, showing that an approximate Lyapunov function for the dynamics exists for memristors with slow decay. Because of the properties of memristors, we have argued that asymptotically the Lyapunov function of a memristive circuit can be written as a quadratic function on spin-like variables, and have connected the coupling matrix to the projector operator on the loop space of the graph \cite{Caravelli2016rl}. 
This result shows a direct connection between purely memristive circuits and Hopfield networks, which we argue minimize a similar type of Lyapunov function. The internally stored patterns are in this case the cycles of the graph. This is interesting for various reasons. 
Insofar it had been argued that a certain degree of external control was necessary in order to perform computation or use memristors as a memory device, either via the use of CMOS or the introduction of capacitors into the network. This paper provides sufficient evidence for the use of memristors in their completely analog regime for computational purposes. This comes at a cost, which is the embedding of the problem of interest in the cycle basis of the graph, which can be a rather nontrivial problem when it is solvable.

The Lyapunov function for the dynamics of memristive circuits can then be used for various purposes. We have first focused on the complexity of these functions in the case of random circuits. We have provided numerical evidence and argued that the couplings scale as in the case of mean-field spin glasses, i.e. the Sherrington-Kirkpatrick model, although only if neglecting correlations between matrix elements of lowest order. Using this approximation we have provided approximate formulae for the number of stationary points, provided some topological considerations. This result explains the importance of the initial conditions for the asymptotic state of the memristors. In fact, it is known (for instance in the case of a glassy system) that for a system which has a large set of local minima there is a large sensitivity to the initial conditions.

As a test-bed of these ideas, we have used the memristor dynamics to see how well (or bad) the dynamics leads to the minimization of a certain quadratic functional 
We have compared the minimization of a combinatorial problem using both simulated annealing and the memristive dynamics introduced in this paper. Although the memristor dynamics could not reach a global minimum, the speed of convergence and the closeness to the Metropolis result suggest the use of these mixed dynamical-combinatorial algorithm to provide quick answers to combinatorial problems on spin-like variables (or alternative 0-1 variables, thus a quadratic unconstrained binary optimization). We do not claim these answers to be optimal, but we find nonetheless interesting that a rather simple procedure like the one we did performed remarkably well compared to simulated annealing.

The main result of this paper however remains the Lyapunov functional for the study of the dynamics of memristors, and to understand the relaxation properties of these circuits (at least when these are controlled with constant voltage). We have shown a direct connection between optimization and memristive circuits, as advocated by other authors, for instance \cite{Traversa2014, pershin13b,pershin11d}. It is inspiring to think that the minimization of the functional depends on the balance between reinforcement-decay properties of the memristive dynamics, along the lines of other similar algorithms \cite{Dorigo} which use collective dynamics as heuristic optimization methods.

Thus, this paper provides needed background work to understand the dynamics of circuits with memory, their sensitivity to initial conditions and their use for computational purposes in the fully analog regime.


\begin{center}
\textbf{Acknowledgement}
\end{center}
We would like to thank J. P. Carbajal, C. Coffrin, M. Vuffray and A. Lokhov for various comments on the results of this paper. In particular, we really appreciated conversations on this same topic with M. Di Ventra, F. L. Traversa and F. Sheldon during my visit at UCSD in 2015. I would like to thank F. Sheldon in particular for pointing out some problems with the original proof on the Lyapunov function.
Also, we acknowledge the support of NNSA for the U.S. DoE at LANL under Contract No. DE-AC52-06NA25396. 

\appendix
\section{Appendix: Properties of $L(W)$.}
\subsection{Derivation of $L(W)$}
The Lyapunov functional provided in the paper is given by:
\begin{eqnarray}
L(\vec W)&=&-\frac{\alpha}{2}\sum_i W_i^2-\frac{\alpha\xi}{3}  \sum_{i} \Omega_{ii}  W_i^3\nonumber \\
&-&\alpha\xi  \sum_{i\neq j} \Omega_{ij}  W_i W_j^2+\frac{1}{\beta} \sum_{ij} W_i \Omega_{ij} S_j.
\label{eq:Lyapunovf}
\end{eqnarray}
In order to derive the functional above, we use the differential equation for the memristive dynamics:
\begin{equation}
(I+\xi \Omega W) \frac{d \vec W}{dt}=(I+\xi \Omega W)\alpha \vec W-\frac{1}{\beta} \Omega \vec S,
\end{equation}
for the case of constant voltages $\vec S$. At this point we can obtain the functional form of the derivative of the Lyapunov function:
\begin{eqnarray}
\frac{d}{dt} L(\vec W)&=&\partial_t L(W)+\sum_i \delta_{w_i}L(W)\frac{d}{dt} W_i\nonumber \\
&=&- \sum_i \frac{d W_i }{d t}[(I+\xi \Omega W)\alpha \vec W-\frac{1}{\beta} \Omega \vec S]_i +\vec M\cdot \frac{d \vec W}{d t}\nonumber \\
&=& -\sum_{ij}\frac{d W_i }{d t}\left(I+\xi \Omega W\right)_{ij}\frac{d W_j }{d t}+ \vec M\cdot \frac{d \vec W}{d t}\nonumber \\
&=&-||\frac{d \vec W}{d t}||^2_{(I+\xi \Omega W)}+ \vec M\cdot \frac{d \vec W}{d t}
\label{eq:lf}
\end{eqnarray}
Where we assumed that $\partial_t L(W)=0$, as $\vec S$ is constant in time. We also introduced
the quantity $M_i=-2 \alpha \xi \sum_{j\neq i} W_j \Omega_{ji} W_i$. The variation of the functional is provided in the subsection below in detail.
First we prove that  $I+\xi \Omega W$ is a positive definite matrix. We use the matrix similarity for eigenvalues,
$I+\xi \Omega W\sim I+\xi \sqrt{W} \Omega \sqrt{W}$
to show that given a vector $\vec b$, its norm is $\vec b^t(I+\xi \sqrt{W} \Omega \sqrt{W})\vec b=||\vec b||^2+\xi ||\sqrt{W}\vec b||_{\Omega} $.
Since $\Omega$ is a projector and is semipositive, then we have that for any non-zero vector $\vec b$, 
$\vec b^t(I+\xi \sqrt{W} \Omega \sqrt{W})\vec b>0.$
We now observe that the first term in eqn. (\ref{eq:lf}) is always negative, meanwhile the second term is always positive, since the inverse of a positive operator is always positive. However, meanwhile the first term does not scale with any parameter, we have that the second term scales roughly as $\alpha^2 \xi^2$. Also, the min norm of the operator $I+\xi \Omega W$ is of order one, which implies that the sup norm of its inverse is also of order one, for arbitrary values of $\xi$.
As a comment, we see that up to this point, this is enough to prove that if $\Omega$ is diagonal the second term disappears, as $\vec M=0$. This completes the proof that $\frac{d}{dt} L(\vec W)\leq 0$ for the case of $\Omega$ diagonal, which is trivial. Also, note that we have $\frac{d}{dt} L(\vec W)=0$ if and only if $\frac{d \vec W}{dt}=0$. This confirms the fact that $L(\vec W)$ is a Lyapunov function in this regime, and thus applies also for weakly interacting memristors.
 
We are interested in the more general case of arbitrary $\Omega$. For this purpose, we consider the following bounds. We have
\begin{equation}
-||\frac{d\vec W}{dt}||^2_{I+\xi \Omega W}+ \frac{d \vec W}{dt} \cdot \vec M\leq -||\frac{d\vec W}{dt}||^2+ ||\frac{d \vec W}{dt}||\cdot ||\vec M||=||\frac{d\vec W}{dt}||\left(||\vec M||- ||\frac{d \vec W}{dt}||\right)
\end{equation}
where we used the Cauchy-Schwarz inequality and the conservative lower bound $||\frac{d\vec W}{dt}||^2_{I+\xi \Omega W}\geq ||\frac{d\vec W}{dt}||^2$. We know already that if $\frac{d}{dt}\vec W=0$, then $\frac{d}{dt} L=0$. Let us focus on $\frac{d}{dt} \vec W\neq 0$. Thus, a conservative bound suggests that we need to bound it via $\text{sup}_{\vec W} ||\vec M||$ and $\text{min}_{\vec W}|| \frac{d}{dt} \vec W||$.
We have shown already that if $\Omega_{ij}$ is diagonal then the proposed Lyapunov function has negative derivative, so we can make a very loose upper bound.
We now have the following  bound on $||\vec M||$:
\begin{equation}
|| \vec M||=2 \alpha \xi \sqrt{\sum_i \sum_{j\neq i} \sum_{j\neq k} W_i^2 \Omega_{ij} W_j \Omega_{ik} W_k}\leq 2 \alpha \xi \sqrt{\sum_{j\neq i, k\neq i} \Omega^2_{jk}} \leq 2\alpha \xi \bar \Omega N,
\end{equation}
where $\bar \Omega=\text{sup}|\Omega_{ij}|$ is the maximum absolute value of the elements of $\Omega_{ij}$.
On the other hand, we have
\begin{eqnarray}
min_{\vec W}|| \frac{d}{dt} \vec W|| &=&min_{\vec W} || \alpha \vec W-\frac{1}{\beta} (I+\xi \Omega W)^{-1}\Omega \vec S|| \nonumber \\
&=&\min_{\vec W} \sqrt{ \alpha^2 || \vec W||^2 +\frac{1}{\beta^2} || (I+\xi \Omega W)^{-1}\Omega \vec S||^2- 2 \frac{\alpha}{\beta} \vec W \cdot (I+\xi \Omega W)^{-1}\Omega \vec S  } \nonumber \\
&=& \sqrt{ \alpha^2 \min_{\vec W}|| \vec W||^2 +\frac{1}{\beta^2}\min_{\vec W}  || (I+\xi \Omega W)^{-1}\Omega \vec S||^2- 2 \frac{\alpha}{\beta}\max_{\vec W} \vec W \cdot (I+\xi \Omega W)^{-1}\Omega \vec S  } \nonumber \\
&=& \sqrt{ \frac{1}{\beta^2}|| (I+\xi \Omega )^{-1}\Omega \vec S||^2- 2 \frac{\alpha}{\beta}\max_{\vec W} \vec W \cdot (I+\xi \Omega W)^{-1}\Omega \vec S  }.
\end{eqnarray}
Now, since $\Omega$ is a projector, we have $ (I+\xi \Omega )^{-1}\Omega=\sum_{j=0}^\infty (-1)^j \xi^j \Omega^{j+1}=\sum_{j=0}^\infty (-1)^j \xi^j \Omega=\frac{1}{1+\xi} \Omega$. Moreover,
\begin{equation}
\max_{\vec W} \vec W \cdot (I+\xi \Omega W)^{-1}\Omega \vec S\leq \frac{N}{(I+\xi )^{-1}} ||\Omega \vec S||
\end{equation}
and thus
\begin{equation}
min_{\vec W}|| \frac{d}{dt} \vec W||=\frac{1}{\sqrt{1+\xi} }\sqrt{\frac{1}{\beta^2}||\Omega \vec S||^2-2\frac{\alpha}{\beta} N||\Omega \vec S||}.
\end{equation}
We thus find that if
\begin{equation}
2\alpha \xi \bar \Omega N-\frac{1}{\sqrt{1+\xi} }\sqrt{\frac{1}{\beta^2}||\Omega \vec S||^2-2\frac{\alpha}{\beta}N ||\Omega \vec S||}<0
\end{equation}
the Lyapunov function we introduced has negative derivative always. Let us now define $|| \Omega \vec S||^2= N^2 s^2(N)$. We can rewrite the inequality as
\begin{equation}
4 \xi^2(1+\xi) \bar \Omega^2 < \frac{s(N)^2}{\alpha^2 \beta^2}-2\frac{s(N)}{\alpha \beta}.
\end{equation}
Because $s(N)\sim O(1)$ in the physical case (e.g. the single voltage element does not scale with the size of the system), the bound above gives a meaningful relationship between the nonlinearity parameter $\xi$ and the mean field dynamical properties of the system, $\frac{s}{\alpha \beta}$, which had been already introduced in \cite{mean-field} in a mean field toy model of memristive dynamics.
It is interesting to note that if we define $\frac{s}{\alpha \beta}=q$, the equation above forms a parabola and an inequality of the form
\begin{equation}
-q^2+2 q+c<0,
\end{equation}
which characterizes the area below the curve. Thus, via the bound we see that below the parabola the system is going into a minimum of the Lyapunov function.

This shows what we had anticipated, e.g. the fact that the Lyapunov function we defined has a negative derivative, and because we have that $\frac{d}{dt} L(W)=0 \leftrightarrow \frac{d}{dt} \vec W=0$ we have the required property for $\xi$ small enough. Also, being $L(W)$ defined on a compact and being smooth, we know it must be bounded from below, which concludes the proof.
Clearly, how proof has a main fallacy: it applies to continuous dynamical systems with continuous derivatives. If certain memristors reach the boundary values, discontinuities appear in the dynamics. Simulations however show that the Lyapunov functional we obtained is an upper bound to the one obtained numerically and including the constraints. 

\subsection{Variation}
Let us now calculate the variation of the functional of eqn. (\ref{eq:lf}). Again, we consider the functional:
\begin{eqnarray}
L(\vec W)&=&-\frac{\alpha}{2}\sum_i W_i^2-\frac{\alpha\xi}{3}  \sum_{i} \Omega_{ii}  W_i^3\nonumber \\
&-&\alpha\xi  \sum_{j\neq i} \Omega_{ji}  W_j W_i^2+\frac{1}{\beta} \sum_{ji} W_j \Omega_{ji} S_i
\end{eqnarray}
We have that:
\begin{eqnarray}
\delta_{W_j} L(W)&=&\delta_{W_j}\Big(-\frac{\alpha}{2}\sum_i W_i^2-\frac{\alpha\xi}{3}  \sum_{i} \Omega_{ii}  W_i^3 \nonumber \\
&-&\alpha\xi  \sum_{i\neq j} \Omega_{ji}  W_j W_i^2+\frac{1}{\beta} \sum_{ji} W_i \Omega_{ji} S_i\Big) \nonumber \\
&=&-\alpha W_j-\alpha\xi \Omega_{jj}  W_j^2 \nonumber \\
&-&\alpha\xi  \sum_{i \left(i \neq j\right)} \Omega_{ji}  W_i^2+\frac{1}{\beta}\Omega_{ji} S_i-2\alpha\xi  \sum_{i \left(i \neq j\right)} W_i \Omega_{ij}  W_j
\end{eqnarray}
which implies 
\begin{eqnarray}
\delta_{\vec W} L(\vec W)&=&-\alpha \vec W-\alpha \xi \Omega \vec W^2+ \frac{1}{\beta} \Omega \vec S-(I+\xi \Omega W) \frac{d}{dt} \vec W+ \vec M\nonumber \\
&=&-\left((I+\xi \Omega W) \alpha \vec W-\frac{1}{\beta} \Omega \vec S\right)-(I+\xi \Omega W) \frac{d}{dt} \vec W+ \vec M\nonumber \\
&=&-(I+\xi \Omega W) \frac{d}{dt} \vec W+\vec M
\end{eqnarray}
where on the last line we used the equations of motion, and defined $M_i= -2\alpha \xi\sum_{j \left(i \neq j\right)} W_j \Omega_{ji}  W_i$. For random networks, 
This mapping is reminiscent of the case of continuous neuronal networks introduced by Little \cite{Little} and then Hopfield in a series of important papers \cite{Hopfield1,Hopfield2}. The Little-Hopfield model has sparked interest from the Statistical Physics community since the very beginning \cite{SGHopfield}.
In the past years this particular line of study has been subject of scrutiny by many experimental groups \cite{HopMemr1,HopMemr2,HopMemr3}. The key difference is that in each case, these were studied in conjunction with ordinary and active (CMOS) electronic components to build Hopfield learning networks. We argue instead that memristive circuits \textit{per se} form a special kind of Hopfield network defined by the Lyapunov function above, without the need of extra components.

\section{Complexity of the Lyapunov functional \textit{via} Kac-Rice formula}
This section provides the details for the average number of stationary points of the Lyapunov functional above in the asymptotic regime. Let us now consider the following average:
\begin{equation}
\langle \# \rangle =\langle\sum_{\{\vec w=1,0\}}  \text{det}(\partial^2_{\partial w_i \partial w_j} L(\vec w)) \prod_i \delta\left(\partial_{w_i} L(\vec w)\right)  \rangle_{P(\Omega)}
\end{equation}
where $L(\vec w)$ is the asymptotic functional in eqn. (\ref{eq:lf}).
We can see that
\begin{eqnarray}
\partial_{\partial w_i} L(\vec w)&=&-\left(\frac{\alpha}{2} +\frac{\alpha\xi}{3}\Omega_{ii} -\frac{1}{\beta} \sum_j \Omega_{ij} S_j\right)  \nonumber \\
&+&\alpha \xi \sum_{j}\left( \Omega_{ij} -\Omega_{ii} \delta_{ij}\right)  w_j
\end{eqnarray}
 and thus
 \begin{equation}
\partial^2_{\partial w_i \partial w_j} L(\vec w)=\alpha \xi \left( \Omega_{ij} -\Omega_{ii} \delta_{ij}\right) 
\end{equation}
We are interested in an asymptotic upper bound on the number of local minima of the function $L(w)$.
We first proceed by calculating the determinant. If $N$ is the number of memristors, then $\text{det}(\alpha \xi \left( \Omega_{ij} -\Omega_{ii} \delta_{ij}\right) )=(\alpha \xi)^N  \text{det}\left( \Omega_{ij} -\Omega_{ii} \delta_{ij}\right) $.
In the scaling observed above, we have noted that $\Omega_{ii}\approx1-\frac{\sqrt{3}}{\sqrt{N}}$, meanwhile $\Omega_{ij}\approx \frac{\sqrt{3}}{\sqrt{N}} Q_{ij}$. Using this scaling, we have: $\text{det}\left( \Omega_{ij} -\Omega_{ii} \delta_{ij}\right)\approx \frac{3^{\frac{N}{2}}}{N^{\frac{N}{2}}}\text{det}\left( Q_{ij}-(\sqrt{N}-\sqrt{3}) \delta_{ij}\right)\approx_{N\gg 1}\text{det}\left( Q_{ij}-\sqrt{N} \delta_{ij}\right)$, with $Q$ a projector which can be written as 
\begin{equation}
Q_{ij}=\sum_{l=1}^L v^l_i {v^l_j}^t .
\end{equation}
We now use the formula:
\begin{equation}
\text{det}(v^l_i {v^l_j}^t+A)=(1+ {v^l}^t A^{-1}v) \text{det}(A) 
\end{equation}
if $A$ is invertible, and write:
\begin{equation}
\text{det}\left( Q_{ij}-\sqrt{N} \delta_{ij}\right)= N^{\frac{N}{2}}\prod_{l=1}^L(1+ {v^l}^t A^{-1}_lv)
\end{equation}
with $A_l=\sum_{k=l+1}^L v^k_i {v^k_j}^t-\sqrt{N} \delta_{ij}$. Since $\sum_{k=l+1}^L v^k_i {v^k_j}^t$ is also a projector on the subspace, we have:
\begin{equation}
A_l^{-1}=\frac{1}{\sqrt{N}(1-\sqrt{N})}\sum_{k=l+1}^L v^k_i {v^k_j}^t-\frac{1}{\sqrt{N}}I.
\end{equation}
It is now easy to see that since $v^l\cdot v^{l^\prime}=0$ if $l\neq l^{\prime}$, we have that
$${v^l}^t A^{-1}_lv=-\frac{1}{\sqrt{N}},$$
and thus we have the remarkable result that in the limit described above, we have
\begin{eqnarray}
\text{det}\left( Q_{ij}-\sqrt{N} \delta_{ij}\right)&=&\frac{3^{\frac{N}{2}}}{N^{\frac{N}{2}}} \left( 1-\frac{1}{\sqrt{N}}\right)^L N^{\frac{N}{2}} \nonumber \\
&=&3^{\frac{N}{2}}  \left( 1-\frac{1}{\sqrt{N}}\right)^L
\end{eqnarray}
which is independent from the distribution of $\Omega$'s elements, but only on the scaling we used. 
We thus obtain a rough scaling of the number of minima, given by:
\begin{equation}
\langle \# \rangle=(\sqrt{3} \alpha \xi)^N \left( 1-\frac{1}{\sqrt{N}}\right)^L Z(\vec S,\Omega)
\label{eq:scaling}
\end{equation}
Up to here no assumption has been made on the distribution of $\Omega_{ij}$, if not the scaling observed numerically.

In order to evaluate $Z(\vec S,\Omega)$, however, assumptions have to be made.
We now focus on evaluating the term:
\begin{eqnarray}
Z(\vec S,P(\Omega)) =\langle\sum_{\{\vec w=1,0\}} \prod_i \delta\left(\partial_{w_i} L(\vec w)\right)  \rangle_{P(\Omega)}
\end{eqnarray}
which can be written as:

\begin{equation}
Z(S,P(\Omega))= \frac{1}{(2\pi)^N}\int \prod_{ij} d\Omega_{ij}\sum_{\{ w\}}\prod_i \int^{\infty}_{-\infty} d\eta_i e^{i \eta_i \left(
-\left(\frac{\alpha}{2} +\frac{\alpha\xi}{3}\Omega_{ii} -\frac{1}{\beta} \sum_j \Omega_{ij} S_j\right)  +\alpha \xi \sum_{j}\left( \Omega_{ij} -\Omega_{ii} \delta_{ij}\right)  w_j\right)} P(\Omega_{ij}).
\end{equation}

We will now introduce again the scaling $\Omega_{ii}=1-\frac{1}{\sqrt{N}}$, and $\Omega_{ij}=\frac{\sqrt{3}}{\sqrt{N}} Q_{ij}$.
We now consider the case in which $P(Q_{ij})=\frac{1}{\sqrt{2\pi \sigma^2}} e^{-\frac{Q_{ij}^2}{2 \sigma^2}}$.
We can use a this point the formula:
\begin{equation}
\int_{-\infty}^\infty dx\ e^{i q x-\frac{x^2}{2b^2}}=\sqrt{2\pi b^2} e^{-\frac{b^2 q^2}{2}}
\end{equation}
and write:

\begin{eqnarray}
Z(S,P(\Omega))&=& \frac{1}{(2\pi)^N}\int \prod_{ij} d\Omega_{ij}\sum_{\{ w\}}\prod_i \int^{\infty}_{-\infty} d\eta_i e^{i \eta_i \left(
-\left(\frac{\alpha}{2} +\frac{\alpha\xi}{3}\Omega_{ii} -\frac{1}{\beta} \sum_j \Omega_{ij} S_j\right)  +\alpha \xi \sum_{j}\left( \Omega_{ij} -\Omega_{ii} \delta_{ij}\right)  w_j\right)} P(\Omega_{ij}) \nonumber \\
&=&\frac{1}{(2\pi)^N}\sum_{\{ w\}}\prod_i \int^{\infty}_{-\infty} d\eta_i e^{i \eta_i \left( \alpha \xi (1-\frac{\sqrt{3}}{\sqrt{N}}) w_j -(\frac{\alpha}{2}+\alpha \xi (1-\frac{\sqrt{3}}{\sqrt{N}})\right)}  \nonumber \\
&\cdot&\int_{-\infty}^{\infty} \prod_{ij} dQ_{ij} e^{i \frac{\sqrt{3}}{\sqrt{N}}\eta_i Q_{ij} \left(\alpha \xi w_j+\frac{1}{\beta} S_j \right)-\frac{Q_{ij}^2}{2 \sigma^2}}\frac{1}{\sqrt{2 \pi \sigma^2}} \nonumber \\
&=&\frac{1}{(2\pi)^N}\sum_{\{ w\}}\prod_i \int^{\infty}_{-\infty} d\eta_i e^{i \eta_i \left( \alpha \xi (1-\frac{\sqrt{3}}{\sqrt{N}}) w_j -(\frac{\alpha}{2}+\alpha \xi (1-\frac{\sqrt{3}}{\sqrt{N}})\right)} e^{-N\frac{\eta_i^2 \sigma^2 \frac{  \left( \alpha \xi w_j+\frac{1}{\beta} S_j  \right)^2 }{3} }{2}} \nonumber
\end{eqnarray}
and if we define $a(w)=\alpha \xi (1-\frac{\sqrt{3}}{\sqrt{N}}) w_j -\left(\frac{\alpha}{2}+\alpha \xi (1-\frac{\sqrt{3}}{\sqrt{N}})\right)$ and $b_i(w)=  \left( \alpha \xi w+\frac{1}{\beta} S_i \right)  $, we can write the integral as:
\begin{eqnarray}
Z(S,P(\Omega))&=&\left(\frac{1}{2\pi}\frac{\sqrt{6 \pi }}{\sqrt{N \sigma ^2}}\right)^N \nonumber \\
&\cdot& \prod_i \left(\frac{ e^{-\frac{3 a(1)^2}{2 N \sigma^2 b_i(1)^2} }}{b_i(1)} +\frac{ e^{-\frac{3 a(0)^2}{2 N \sigma ^2 b_i(0)^2}}}{b_i(0)}\right)
\end{eqnarray}
We now rescale $\sigma^2\rightarrow \sigma^2/N$ in order for $\sigma$ to be physical in the limit $N\rightarrow \infty$.
We thus obtain, in the approximation in which $S_i=S$, $Z(S)$ becomes:
\begin{equation}
Z(S)=\left(\frac{\sqrt{3 }}{\sqrt{ \pi \sigma ^2}}\right)^N\left(\frac{ e^{-\frac{3 a(1)^2}{2  \sigma^2 b(1)^2} }}{b(1)} +\frac{ e^{-\frac{3 a(0)^2}{2  \sigma ^2 b(0)^2}}}{b(0)}\right)^N.
\end{equation}
In the limit $\alpha \xi\gg S\gg1$, we have $\frac{ e^{-\frac{3 a(1)^2}{2  \sigma^2 b(1)^2} }}{b(1)} +\frac{ e^{-\frac{3 a(0)^2}{2  \sigma ^2 b(0)^2}}}{b(0)}\approx \frac{1}{\alpha \xi}$. Thus, we have the final formula:
\begin{equation}
\langle \# \rangle\approx \left(1-\frac{1}{\sqrt{N}}\right)^L (\frac{ \sqrt{3} }{\sqrt{\pi}\sigma} )^N.
\end{equation}
which shows two regimes. If $\sigma<\sqrt{3/ \pi}\approx 0.977205$, the number of stationary points increases exponentially.



\begin{thebibliography}{10}

\bibitem{indiveri}
G. Indiveri,    S.-C. Liu,  Memory and information processing in neuromorphic systems, Proceedings of IEEE, 103:(8) 1379-1397 (2015)


\bibitem{Avizienis}
A.V. Avizienis et al., Neuromorphic Atomic Switch Networks. PLoS ONE 7(8): e42772. (2012)

 \bibitem{Stieg12} A. Z. Stieg, A. V. Avizienis et al., Emergent Criticality in Complex Turing B-Type Atomic Switch Networks, Adv. Mater., 24: 286-293  (2012)



\bibitem{traversa14b}
F. L. Traversa, M.~Di~Ventra.
\newblock Universal memcomputing machines, \newblock { IEEE Trans. Neural Netw. Learn. Syst., (DOI:
  10.1109/TNNLS.2015.2391182, preprint arXiv:1405.0931)} 

\bibitem{QUBO} F. Glover, G. Kochenberger, Gary, A Tutorial on Formulating and Using QUBO Models,. arXiv:1811.11538 (2019)

\bibitem{Caravelli2015}
F. Caravelli, A. Hamma, M. Di Ventra, Scale-free networks as an epiphenomenon of memory, EPL, 109, 2 (2015)


 \bibitem{Caravelli2016}
F. Caravelli, Trajectories entropy in dynamical graphs with memory, Front. Robot. AI 3, 18 (2016), arXiv:1511.07135

\bibitem{Caravelli2019} F. Caravelli, J. P. Carbajal, Memristors for the curious outsiders, Technologies 2018, 6(4), 118
\bibitem{Caravelli20192} A. Zegarac, F. Caravelli, Memristive networks: From graph theory to statistical physics ,2019 EPL Perspectives 125 10001



\bibitem{Caravelli2016rl} F. Caravelli, F. L. Traversa, M. Di Ventra, The complex dynamics of memristive circuits: analytical results and universal slow relaxation, Phys. Rev. E 95, 2 (2017) 

\bibitem{strukov08} D.B. Strukov, G. Snider, D.R. Stewart, and R.S. Williams, The missing memristor found, Nature 453, pp. 80-83 (2008)

\bibitem{chua76a}
L.~O. Chua, S.~M. Kang.
\newblock Memristive devices and systems,
\newblock { Proc. IEEE}, 64:209--223, 1976.

\bibitem{stru12}
J. J. Yang, D. B. Strukov, D. R. Stewart, Memristive devices for computing, Nature Nanotechnology 8 (2013)

\bibitem{AtomicSwitch1}
Ohno et al., Sensory and short-term memory formations observed in a Ag2S gap-type atomic
switch, Appl. Phys. Lett. 99, 203108 (2011);
\bibitem{AtomicSwitch2} Wang et al, Memristors with diffusive dynamics as synaptic emulators for neuromorphic computing Nat. Mat. 16 (2017)

\bibitem{mean-field} F. Caravelli, P. Barucca, Exactly solvable model of memristive circuits: Lyapunov function and mean field theory, \href{arXiv:1706.03001}{arXiv:1706.03001} EPL 122(4) 2018


\bibitem{Caravelli2017} F. Caravelli, Locality of interactions for planar memristive circuits, Phys. Rev. E 96, 052206 (2017)

\bibitem{Caravelli2016ml} F. Caravelli,  Advances in Memristive Networks, ed. Adamatzky et al., Int. J. of Par. and Dist. Sys. 0 (2017),  arXiv:1611.02104





\bibitem{Little} W. A. Little,   The existence of persistent states in the brain, Math. Biosci., 19, 101-120 (1974).
\bibitem{Hopfield1} J. J. Hopfield, "Neural networks and physical systems with emergent collective computational abilities", Proceedings of the National Academy of Sciences of the USA, vol. 79 no. 8 pp. 2554-2558, (1982)
\bibitem{Hopfield2} J. J. Hopfield, D. W. Tank, Computing with neural circuits: a model, Science  233, Issue 4764, pp. 625-633 (1986)
\bibitem{SGHopfield} D. J. Amit, H. Gutfreund, H. Sompolinsky, Spin-glass models of neural networks, Phys. Rev. A 32, 2 (1985)
\bibitem{HopMemr1} S.G. Hu et al., Associative memory realized by a reconfigurable memristive Hopfield neural network, Nat. Comm. 6,  7522 (2015)
\bibitem{HopMemr2} S. Kumar et al, Chaotic dynamics in nanoscale NbO2 Mott memristors for analogue computing, Nature 548, 318?321 (2017)
\bibitem{HopMemr3} M. Tarkov, Hopfield Network with Interneuronal Connections Based on Memristor Bridges, Advances in Neural Networks ? ISNN 2016 pp 196-203 (2016)

\bibitem{rgf}  R.J. Adler, J. E. Taylor, Random Fields and Geometry, Springer-Verlag, 
New York, (2007)

\bibitem{fyodorov} Y. V. Fyodorov, High-Dimensional Random Field and Random Matrix Theory, Markov Processes Relat. Fields 21, 483--518 (2015)

\bibitem{parisi} G., Parisi, A sequence of approximate solutions to the S-K model for spin glasses, J. Phys.
A 13, L-115 (1980)

\bibitem{poloni} F. Poloni, Algorithms for Quadratic Matrix and Vector Equations, Theses (Scuola Normale Superiore), Pisa (2011)

\bibitem{Traversa2014}
F. L. Traversa, C. Ramella, F. Bonani, M. Di Ventra, Memcomputing NP-complete problems in polynomial time using polynomial resources and collective states, Science Advances, vo. 1, no. 6, pag e1500031 (2015)



\bibitem{pershin11d}
Y.~V. Pershin, M. Di~Ventra.
\newblock Solving mazes with memristors: a massively-parallel approach,  \newblock { Phys. Rev. E}, 84:046703 (2011)

\bibitem{pershin13b}
Y.~V. Pershin, M. Di~Ventra.
\newblock Self-organization and solution of shortest-path optimization problems
  with memristive networks, \newblock { Phys. Rev. E}, 88:013305 (2013)

\bibitem{Dorigo} M. Dorigo, L. M. Gambardella, Ant colonies for the traveling salesman problem, Biosystems
 43,  2, p 73-81 (1997)
 \bibitem{cycle} T. Kavitha et al., Cycle bases in graphs characterization, algorithms, complexity, and applications, Computer Science Review 3 (4), pp 199-243 (2009)

\bibitem{heur} T.-J. Chang et al., Heuristics for cardinality constrained portfolio optimisation, Computers \& Operations Research 27  1271-1302 (2000)
\bibitem{heur2} 225 Nikkei asset dataset,  OR-Library, http://people.brunel.ac.uk/~mastjjb/jeb/orlib/files/port5.txt
\bibitem{Markowitz} H. Markowitz, Portfolio Selection, J. of Fin. 7, 1 (1952), pp. 77-91
\bibitem{simann} S. Kirkpatrick, C. D. Gelatt Jr., M. P. Vecchi, Optimization by Simulated Annealing, Science  Vol. 220, Issue 4598, pp. 671-680 (1983)





\end{thebibliography}
\end{document}